\documentclass[
  reprint,
  aps,
  prl,
  superscriptaddress,
  amsmath,
  amssymb,
  longbibliography
]{revtex4-2}

\usepackage{graphicx}
\usepackage{dcolumn}
\usepackage{bm}
\usepackage[colorlinks=true,citecolor=blue,urlcolor=blue,linkcolor=blue]{hyperref}
\usepackage{CJK}

\begin{document}
\begin{CJK*}{UTF8}{}
\CJKfamily{min}
\title{Funnel-like protein energy landscapes emerge from functional evolution under thermal fluctuations}

\author{Norifumi Maruyama (丸山恭史)}
\affiliation{Department of Physics, The University of Osaka}

\author{Macoto Kikuchi (菊池誠)}
\affiliation{D3 Center, The University of Osaka}

\date{\today}

\begin{abstract}
Foldable proteins exhibit funnel-like energy landscapes, but their evolution
under selection for function remains unclear.  We study this in a lattice
protein model, defining fitness as the equilibrium probability of a fixed
active-site motif and using multicanonical sequence sampling.  At intermediate
temperature, rare high-fitness sequences show funnel-like energy and double-well
free-energy landscapes.  At very low temperature, abundant high-fitness
sequences retain glass-like landscapes.  Thus, thermal fluctuations turn a
local functional requirement into a global structural constraint.
\end{abstract}

\maketitle
\end{CJK*}

\section*{Introduction}

Proteins function by folding into specific native structures under
physiological conditions; according to Anfinsen's thermodynamic hypothesis, a
protein's native structure is the thermodynamically stable equilibrium state
determined by its amino-acid sequence~\cite{Epstein1963,Anfinsen1973}.
G{\=o}'s consistency principle states that local and nonlocal interactions in
a foldable protein are mutually consistent in stabilizing its native
structure~\cite{Go1983}.  Building on this principle, Bryngelson and Wolynes
incorporated concepts from spin-glass theory into a statistical-mechanical
theory of protein folding and formulated the principle of minimal frustration:
evolution suppresses
energetic conflicts from competing conformations, thereby creating a global
bias toward the native state~\cite{Bryngelson1987}.

This native-directed bias is a central ingredient of the modern folding-funnel
picture~\cite{Bryngelson1995,Dill1997}.  In a funnel-like
landscape, the energy
decreases overall as conformations approach the native structure, while the
number of accessible conformations decreases sharply.  The resulting balance
between energy and conformational entropy produces folded and denatured basins
in the free-energy landscape separated by a barrier; as temperature is lowered
through the folding temperature, the native basin becomes thermodynamically
favored, yielding a cooperative two-state transition.  These principles
explain how natural proteins can fold, but leave a fundamental evolutionary
question.  Our hypothesis is that natural selection acts primarily on
biological function rather than directly on foldability.  How, then, can
functional selection generate the global energetic organization required for
folding?

Saito, Sasai, and Yomo first addressed this question by performing evolutionary
simulations of a spin-glass-like protein model under selection for a prescribed local
active-site configuration~\cite{Saito1997}.  They found that, as fitness
increased during evolution, the trajectory-based conformational entropy
decreased and a single functional spin configuration came to dominate, with a
growing energy gap separating this lowest-energy configuration from the bulk
of the spectrum.  The convergence of folding trajectories and site-dependent
frustration further indicated global folding ability and an anisotropic funnel,
without direct selection for a native fold.
Subsequent studies with coarse-grained chain models further supported
functional selection as a route to protein-like global organization and folding
ability~\cite{Yomo1999,Sasaki2002,Nagao2005}.  Whether functional
selection produces a funnel-like energy landscape characterized directly in
terms of explicit chain conformations and the associated two-state
thermodynamic behavior, however, remained unresolved.

The role of thermal fluctuations was isolated by Sakata, Hukushima, and Kaneko
in an evolutionary Ising-spin model~\cite{Sakata2009}.  They found that
evolution under an intermediate level of thermal noise suppresses frustration
around target spins and produces local-Mattis, funnel-like relaxation that is
robust to both thermal noise and mutation, whereas evolution at low temperature
retains spin-glass-like dynamics.  This result identified thermal noise as a
control parameter for the evolution of funnel-like organization.  Because
their spin model contains no geometrical chain conformations, however, it
cannot address the conformational energy landscape of a protein chain.

Here we ask whether functional selection generates a funnel-like energy
landscape in an explicit-chain protein model.  We further ask how its emergence
depends on environmental temperature.  We address these questions by studying
a two-dimensional lattice protein model at equilibrium, with fitness defined as
the probability that a prescribed local active-site motif is realized at
environmental temperature $T$; neither a
native conformation nor foldability enters the fitness.  Enumeration of the
conformational space allows us to construct the energy and free-energy
landscapes directly.  Multicanonical sampling in
sequence space then gives representative sequences across the full fitness
range.  We show that, at an appropriate environmental temperature, selection
to maintain a local functional motif under thermal fluctuations necessarily
generates a funnel-like energy landscape and two-state thermodynamics.  This
provides an underlying thermodynamic mechanism linking functional evolution
to energetic consistency and minimal frustration.
\section*{Model and Methods}

\begin{figure}[t]
\includegraphics[width=\columnwidth]{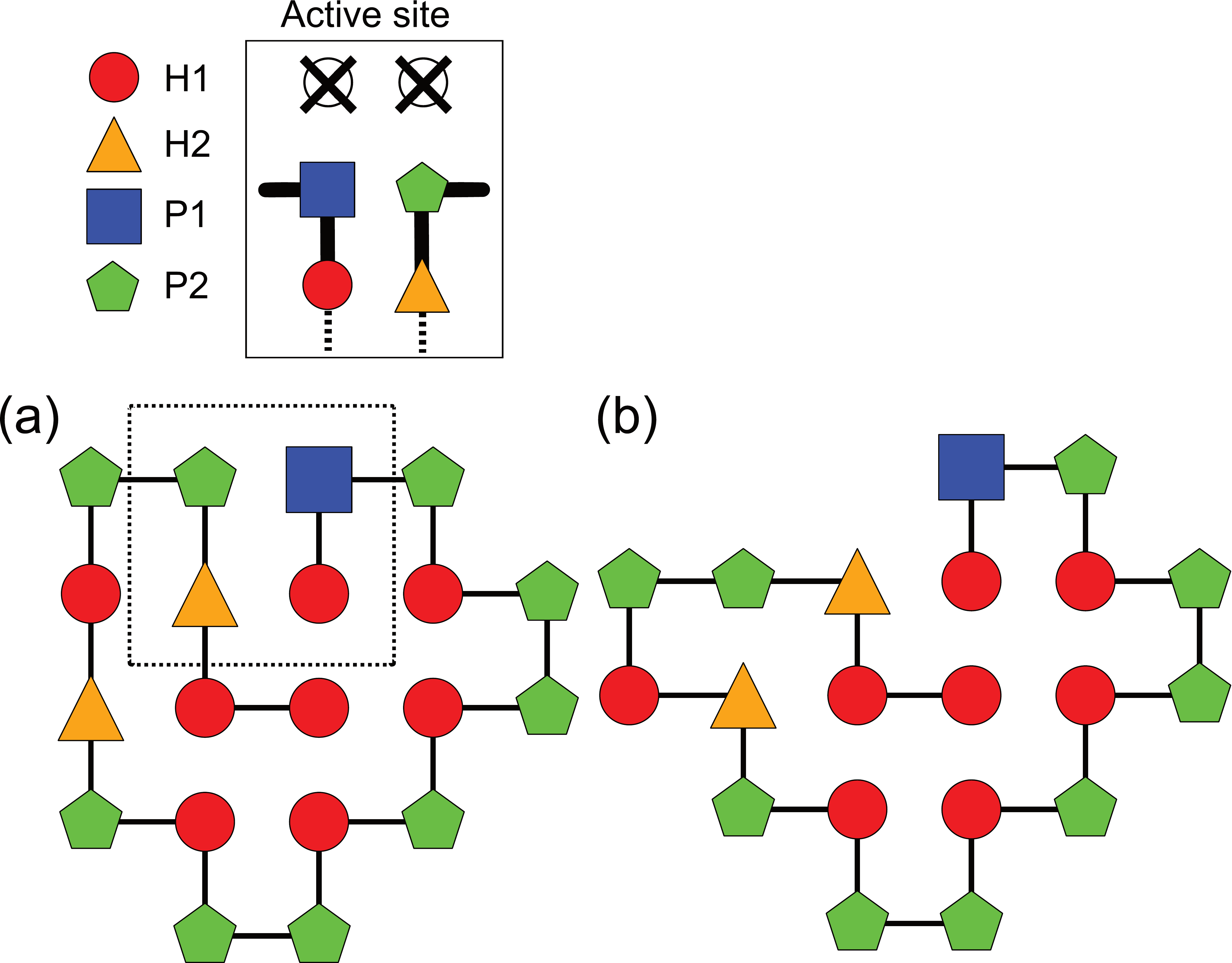}
\caption{Lattice protein and prescribed active site.  Residue types are
distinguished by symbol and color.  The active site comprises two exposed polar
residues supported by a hydrophobic pair; crossed circles denote lattice sites
that must be empty.  Conformation (a) contains the active site (dotted box),
whereas conformation (b) does not.}
\label{FigModel}
\end{figure}

We represent a protein by a self-avoiding chain of 20 residues on a square
lattice [Fig.~\ref{FigModel}].  Motivated by experimental and computational
evidence that five-letter alphabets can retain protein-like folding
properties~\cite{Riddle1997,Wang1999}, we use a slightly simpler four-letter
alphabet comprising two hydrophobic ($\mathrm{H1},\mathrm{H2}$) and two
oppositely charged polar ($\mathrm{P1},\mathrm{P2}$) residue types.  The
commonly used two-letter hydrophobic--polar model is too simple for the present
landscape analysis and suffers from extensive accidental degeneracy.  Our
four-letter representation retains its basic hydrophobic--polar distinction
while avoiding these limitations (see End Matter).  Let $a_i$ denote the
residue type at chain position $i$.  Nonbonded nearest-neighbor residue pairs
interact through $\epsilon(a_i,a_j)$, given in
Table~\ref{tab:contact_energy}; the matrix
preserves the qualitative hierarchy of the Miyazawa--Jernigan contact
potential~\cite{Miyazawa1999}.  The energy of conformation $C$ is

\begin{equation}
E(C)=\sum_{\langle i,j\rangle}\epsilon(a_i,a_j),
\label{eq:energy}
\end{equation}
where the sum runs over all noncovalent nearest-neighbor residue pairs.

\begin{table}[t]
\caption{Contact energies $\epsilon(a,b)$ for the four residue types.}
\label{tab:contact_energy}
\centering
\begin{ruledtabular}
\begin{tabular}{c|rrrr}
 & $\mathrm{H1}$ & $\mathrm{H2}$ & $\mathrm{P1}$ & $\mathrm{P2}$ \\
\hline
$\mathrm{H1}$ & $-3.5$ & $-3.1$ & $-1.2$ & $-1.1$ \\
$\mathrm{H2}$ & $-3.1$ & $-2.3$ & $-0.9$ & $-0.8$ \\
$\mathrm{P1}$ & $-1.2$ & $-0.9$ & $-0.2$ & $-0.4$ \\
$\mathrm{P2}$ & $-1.1$ & $-0.8$ & $-0.4$ & $\phantom{-}0.0$
\end{tabular}
\end{ruledtabular}
\end{table}

We represent biological function by a prescribed local motif that mimics an
active-site geometry: two exposed polar residues supported by a hydrophobic pair
[Fig.~\ref{FigModel}].  Let
$\phi(C)=1$ if this motif occurs anywhere in $C$, and $0$ otherwise.  For a
sequence, fitness is its equilibrium
probability of realizing the motif at environmental temperature $T$,

\begin{equation}
f(T)=\frac{\sum_C\phi(C)e^{-E(C)/T}}
{\sum_C e^{-E(C)/T}}.
\label{eq:fitness}
\end{equation}
Here we set $k_{\rm B}=1$, so $T$ is measured in the same units as the contact
energies.  Thus selection acts only on a local functional structure under thermal
fluctuations.  No native conformation or measure of foldability is specified.
In particular, evaluating $f(T)$ requires no choice
of a native state.

Only for the subsequent landscape analysis do we identify the ground-state
conformation $C_{\rm N}$ of each analyzed sequence as its native conformation;
all high-fitness sequences used for this analysis have a nondegenerate ground
state.  Its native-contact set
$\mathcal{Q}_{\rm N}$ consists of all noncovalent residue pairs $(i,j)$ that
occupy nearest-neighbor lattice sites in $C_{\rm N}$.  The number of native
contacts in any other conformation $C$ is then

\begin{equation}
N_{\rm nc}(C)=\sum_{(i,j)\in\mathcal{Q}_{\rm N}}
\Delta_{ij}(C),
\label{eq:native_contacts}
\end{equation}
where $\Delta_{ij}(C)=1$ when residues $i$ and $j$ form the same noncovalent
nearest-neighbor contact in $C$, and $0$ otherwise.  Thus $N_{\rm nc}$ serves as
a structural order parameter measuring similarity to the native conformation:
larger $N_{\rm nc}$ means that more native contacts have formed.
Taking $N_{\rm nc}$ as the reaction coordinate and denoting its value by $n$,
the equilibrium free-energy landscape is

\begin{equation}
F(n;T)=-T\ln\!\left[
\sum_{C:N_{\rm nc}(C)=n}e^{-E(C)/T}
\right]
\label{eq:free_energy}
\end{equation}

At each $T$, we perform multicanonical Monte Carlo sampling in sequence space,
using fitness $f$ as the sampling variable to obtain sequences approximately
uniformly over its entire accessible range~\cite{Saito2013,Nagata2020,
Kaneko2022,Kikuchi2024}.  Here sequences are the microscopic states, whereas
fitness is the variable whose distribution is flattened, playing the role of
energy in conventional multicanonical sampling~\cite{Berg1992a,Berg1992b}.  Unlike the energy of a
microscopic state, however, the fitness of a sequence is an equilibrium
probability and must therefore be evaluated from the conformational sum in
Eq.~(\ref{eq:fitness}).  The interval
$0\le f\le1$ is divided into 100 bins.  We use the entropic-sampling method,
a variant of multicanonical Monte
Carlo~\cite{Lee1993}.  The
required multicanonical weight is determined beforehand by the Wang--Landau
method~\cite{Wang2001a,Wang2001b}.  To reduce the computational cost of
repeated fitness evaluations, we restrict the sums in Eq.~(\ref{eq:fitness})
to the complete set of conformations with 9--12 noncovalent contacts.  These compact
conformations contain the thermally relevant low-energy states; within this
restricted ensemble, fitness is therefore evaluated by an exact equilibrium
sum rather than by conformational sampling.  The effect of this restriction is
examined for the highest-fitness sequence at $T=1.0$ using the complete
conformational ensemble in the Supplemental Material.  Further details of
sampling are given in the End Matter.
\section*{Results}

\begin{figure}[t]
\includegraphics[width=\columnwidth]{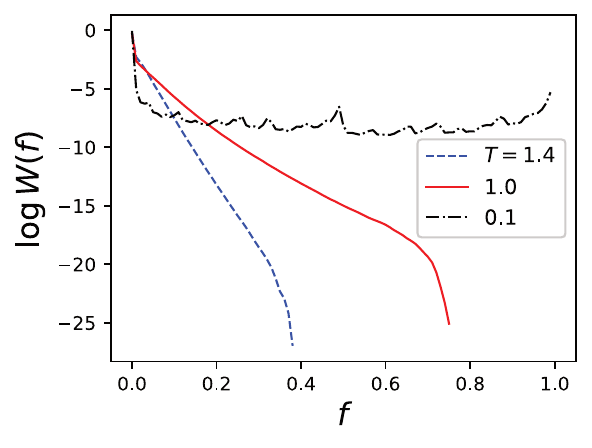}
\caption{Genotypic entropy versus fitness at environmental temperatures
$T=0.1$, $1.0$, and $1.4$.  Each sequence density is normalized over all
fitness bins.}
\label{FigEntropy}
\end{figure}

Figure~\ref{FigEntropy} shows the genotypic entropy $\log W(f)$, where
$W(f)$ is the density of sequences (genotypes) at fitness $f$.  It thereby
quantifies how rare high-fitness sequences are at each temperature.  At $T=1.0$, it decreases rapidly with
$f$, showing that highly functional sequences are very rare; the maximum
fitness is $f_{\max}\simeq0.758$.  At $T=0.1$, by contrast, it remains nearly constant: high
fitness can be achieved by a large variety of sequences.  At $T=1.4$ the
distribution terminates near $f\simeq0.4$, so thermal fluctuations are too
strong for any sequence to maintain the active-site motif with high
probability.  In the subsequent landscape analysis, we therefore focus on
high-fitness sequences at $T=1.0$ and $T=0.1$.

\begin{figure*}[t]
\includegraphics[width=\textwidth]{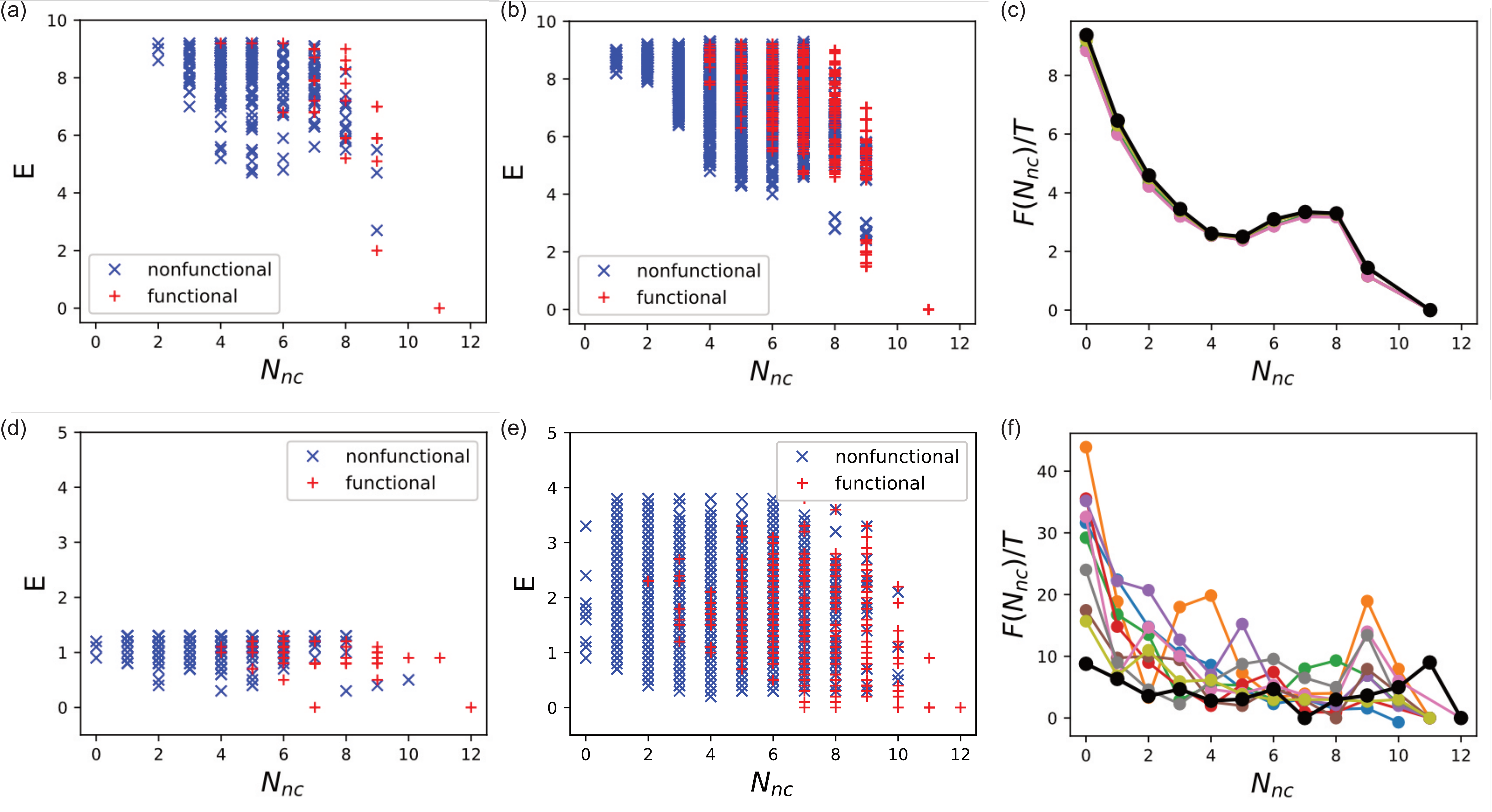}
\caption{Energy and free-energy landscapes of high-fitness sequences obtained
with fitness defined at $T=1.0$ (upper row) and $T=0.1$ (lower row).
(a) Energy spectrum of the highest-fitness sequence at $T=1.0$; (b) spectra of
the 100 highest-fitness sequences; and (c) dimensionless free-energy landscapes
of the ten highest-fitness sequences.  Panels (d)--(f) show the corresponding
quantities at $T=0.1$: (d) energy spectrum of the highest-fitness sequence
satisfying $f<0.9$; (e) spectra of the ten highest-fitness sequences in this
regime; and (f) dimensionless free-energy landscapes of the same ten sequences.
Panels (a), (b), (d), and (e) show the 400 lowest-energy conformations
per sequence; $E$ is measured from each ground-state energy, and $N_{\rm nc}$
is defined relative to each sequence's native conformation.  Red plus signs
and blue crosses denote functional and nonfunctional conformations, respectively.  In (c)
and (f), the additive constant is chosen separately for each sequence so that
$\min_{N_{\rm nc}}F(N_{\rm nc})=0$; the black curves correspond to the
sequences in (a) and (d), respectively, and colors distinguish the remaining
sequences only.}
\label{FigLandscape}
\end{figure*}

The two temperatures produce qualitatively different energy and free-energy
landscapes when resolved by the number of native contacts $N_{\rm nc}$
(Fig.~\ref{FigLandscape}).  The native conformation of the
highest-fitness sequence at $T=1.0$ is shown in Fig.~\ref{FigModel}(a).  Its low-energy
spectrum descends overall as the number of native contacts
$N_{\rm nc}$ increases, with functional conformations concentrated in the low-energy,
native-like region [Fig.~\ref{FigLandscape}(a)].  The same organization occurs
across the 100 highest-fitness sequences [Fig.~\ref{FigLandscape}(b)].  Because
$N_{\rm nc}$ measures structural similarity to the native conformation, this
reproducible native-directed descent characterizes a funnel-like energy
landscape.  The dimensionless free-energy
landscapes $F(N_{\rm nc})/T$ of the ten
highest-fitness sequences nearly collapse, each exhibiting native and
denatured minima separated by a barrier, the thermodynamic signature of
two-state folding [Fig.~\ref{FigLandscape}(c)].  The black curve corresponds
to the highest-fitness sequence in
Fig.~\ref{FigLandscape}(a).  For this sequence, taking $N_{\rm nc}=6$--8 as
the intervening barrier region, we define $T_f$ by equal equilibrium
probabilities of the denatured and native ensembles.  The complete
conformational ensemble gives $T_f\simeq1.23$, so the environmental temperature
$T=1.0$ lies below its folding temperature (Supplemental Material).

Since evolution proceeds through stochastic mutation and selection rather
than deterministic optimization, our analysis at $T=0.1$ focuses on the
broadly populated, sufficiently high-fitness regime rather than only on the
extreme $f\simeq1$ tail.  We
characterize this regime using the highest-fitness sequences with $f<0.9$.
The low-energy spectrum of the highest-fitness sequence shows no systematic
decrease with $N_{\rm nc}$ [Fig.~\ref{FigLandscape}(d)], while those of the ten
highest-fitness sequences vary strongly and share no native-directed trend
[Fig.~\ref{FigLandscape}(e)].  Their free-energy
landscapes are rugged and lack a reproducible two-basin structure
[Fig.~\ref{FigLandscape}(f)].  Surveys of 200 sequences confirm that these
free-energy landscapes remain strongly heterogeneous, in sharp contrast to the
broadly reproducible two-basin organization at $T=1.0$ (Supplemental
Material).  For sequences near the low-temperature
fitness maximum ($f\simeq1$), the free-energy landscapes decline overall with
increasing $N_{\rm nc}$ but remain rugged and strongly sequence dependent
(Supplemental Material).

The temperature contrast extends to global structure.  After chain-reversed
pairs are identified, 500 high-fitness sequence patterns at $T=1.0$ are
classified into only four native-structure classes.  In a survey of 200
high-fitness sequences, their
free-energy landscapes likewise group by these four native-structure classes,
with nearly identical landscape shapes within each class
(Supplemental Material).  By contrast, at $T=0.1$ the 200 highest-fitness
sequences with $f<0.9$ span 187 distinct native structures, and the
extreme-fitness set spans 107 after the same chain-reversal identification
(Supplemental Material).  Local
functional selection therefore strongly restricts both the energy landscape
and fold space at intermediate temperature, whereas low-temperature
high-fitness sequences remain glass-like and structurally diverse.

These conclusions are robust to the computational procedures.  For the
highest-fitness sequence at $T=1.0$, replacing the restricted
9--12-contact ensemble with the complete conformational ensemble produces an
almost identical free-energy landscape (Supplemental Material).
We also performed explicit population-based evolutionary simulations with
fitness evaluated at $T=1.0$.  In two of 40 independent runs, evolution
reached the highest-fitness sequence found by multicanonical sampling, showing
that it is accessible through ordinary mutation and selection (Supplemental
Material).
\section*{Discussion}

Our central result is that selection for a prescribed active site under
appreciable thermal fluctuations generates globally funnel-like energy
landscapes and two-state thermodynamics, even though fitness is defined without
reference to either a native conformation or foldability.  At low temperature, by contrast, high
fitness generally remains compatible with rugged landscapes and structural
diversity.

The temperature dependence clarifies the underlying physical mechanism.  Within
an intermediate temperature range, high fitness requires the active-site geometry to be
preserved not only in a ground state but across the thermally populated
low-energy ensemble.  Sequences that stabilize competing low-energy
conformations lacking the active-site geometry have lower fitness and are
therefore selected against.
This selection pressure thereby constrains global structure: it favors sequences
for which increasingly native-like conformations become energetically
preferred and restricts the high-fitness ensemble to a small set of folds.
At low temperature, by contrast, thermally excited conformations carry little
equilibrium weight.  A sequence can then achieve high fitness
whenever one of its lowest-energy conformations contains the active site,
without organizing the rest of its energy spectrum.  Thermal fluctuations thus
convert a local functional constraint into a global constraint on the energy
landscape.  Although folding kinetics
were not simulated, the rugged landscapes of these sequences are expected to
impede equilibration, so their equilibrium fitness may not be realized on
folding timescales.  We therefore did not perform evolutionary simulations at
low temperature.

Our results provide a direct equilibrium realization of the
functional-selection scenario introduced by Saito, Sasai, and
Yomo~\cite{Saito1997} and developed in subsequent
studies~\cite{Yomo1999,Sasaki2002,Nagao2005}: explicit conformational
enumeration resolves funnel organization in the energy spectrum along the
structural coordinate $N_{\rm nc}$ and the resulting two-state free-energy
structure in $F(N_{\rm nc})$.

This temperature-dependent contrast is consistent with Sakata, Hukushima, and
Kaneko, who found funnel-like relaxation under intermediate thermal noise but
spin-glass-like dynamics at low temperature in an evolutionary Ising-spin
model~\cite{Sakata2009}.  Our explicit-chain model resolves the corresponding
organization directly in the conformational energy spectrum and free-energy
landscape.

Our results further suggest a concrete evolutionary basis for G{\=o}'s
consistency principle~\cite{Go1983} and the principle of minimal frustration
formulated by Wolynes and coworkers~\cite{Bryngelson1987,Bryngelson1995}, as
well as for the folding-funnel picture summarized by Dill and
Chan~\cite{Dill1997}.  At an appropriate
environmental temperature, selection to maintain a local functional geometry
disfavors sequences whose interactions stabilize competing nonfunctional
conformations, thereby making interactions throughout the chain consistent
with the functional fold and suppressing frustration.  Thus, evolution need
not select directly for either consistency or minimal frustration; at an
appropriate environmental temperature, funnel-like organization follows as a
thermodynamic consequence of selection for protein function.

The same mechanism may bear on two longstanding puzzles: why a small active
site is embedded in a much larger protein scaffold~\cite{Srere1984}, and why
the repertoire of protein folds is so limited~\cite{Chothia1992}.  The
requirement that a local functional geometry remain stable at an appropriate
environmental temperature constrains interactions throughout the chain,
providing a thermodynamic rationale for the surrounding scaffold.  That many
high-fitness sequences encode only a few native structures further suggests
that functional requirements restrict the accessible fold repertoire.

Our formulation also connects directly to protein design.  Under the maximum-target-
probability (MTP) criterion, a sequence is designed by maximizing the
equilibrium probability of a prescribed native conformation~\cite{Deutch1996,
Iba1998,Takahashi2021}.  This criterion is recovered from
Eq.~(\ref{eq:fitness}) when $\phi(C)$ selects a single target conformation.
From an evolutionary perspective, our results instead suggest a more natural
design principle: maximize the equilibrium probability of a local functional
structure rather than that of a prescribed global conformation.  With
$\phi(C)$ selecting any conformation containing the local functional motif,
neither the native fold nor its scaffold is specified in advance; both emerge
from selection for function under thermal fluctuations.

The proposed mechanism is most naturally relevant to the early evolution of
small, independently folding globular proteins, before the emergence of more
elaborate multidomain architectures.  The deliberately minimal model
nevertheless captures the essential physics: equilibrium selection on a
thermally maintained local structure produces the temperature-dependent
contrast between funnel-like organization and glass-like behavior.
The principal limitations are the short, two-dimensional chain, which lacks
realistic secondary structure, and the four-letter alphabet.  Future work
should test the mechanism in longer three-dimensional models with richer
alphabets and with different active-site motifs.

In conclusion, at an appropriate environmental temperature, selection to
stabilize a local functional motif generates a funnel-like energy landscape
and two-state thermodynamics without direct selection for foldability.  At low
temperature, the same local requirement remains compatible with glass-like
landscapes and extensive structural diversity.  More broadly, our results show
how thermal fluctuations transform a local functional requirement into global
organization of the conformational energy landscape through evolutionary
selection.
\begin{acknowledgments}
The authors thank Hajime Yoshino, Masaki Sasai, Shuji Ishihara, and Kota Mitsumoto for fruitful discussions and comments. This work was supported by JSPS KAKENHI Grant Numbers JP23K03261 and JP26K06963. 
\end{acknowledgments}


\begin{thebibliography}{29}%
\makeatletter
\providecommand \@ifxundefined [1]{%
 \@ifx{#1\undefined}
}%
\providecommand \@ifnum [1]{%
 \ifnum #1\expandafter \@firstoftwo
 \else \expandafter \@secondoftwo
 \fi
}%
\providecommand \@ifx [1]{%
 \ifx #1\expandafter \@firstoftwo
 \else \expandafter \@secondoftwo
 \fi
}%
\providecommand \natexlab [1]{#1}%
\providecommand \enquote  [1]{``#1''}%
\providecommand \bibnamefont  [1]{#1}%
\providecommand \bibfnamefont [1]{#1}%
\providecommand \citenamefont [1]{#1}%
\providecommand \href@noop [0]{\@secondoftwo}%
\providecommand \href [0]{\begingroup \@sanitize@url \@href}%
\providecommand \@href[1]{\@@startlink{#1}\@@href}%
\providecommand \@@href[1]{\endgroup#1\@@endlink}%
\providecommand \@sanitize@url [0]{\catcode `\\12\catcode `\$12\catcode
  `\&12\catcode `\#12\catcode `\^12\catcode `\_12\catcode `\%12\relax}%
\providecommand \@@startlink[1]{}%
\providecommand \@@endlink[0]{}%
\providecommand \url  [0]{\begingroup\@sanitize@url \@url }%
\providecommand \@url [1]{\endgroup\@href {#1}{\urlprefix }}%
\providecommand \urlprefix  [0]{URL }%
\providecommand \Eprint [0]{\href }%
\providecommand \doibase [0]{https://doi.org/}%
\providecommand \selectlanguage [0]{\@gobble}%
\providecommand \bibinfo  [0]{\@secondoftwo}%
\providecommand \bibfield  [0]{\@secondoftwo}%
\providecommand \translation [1]{[#1]}%
\providecommand \BibitemOpen [0]{}%
\providecommand \bibitemStop [0]{}%
\providecommand \bibitemNoStop [0]{.\EOS\space}%
\providecommand \EOS [0]{\spacefactor3000\relax}%
\providecommand \BibitemShut  [1]{\csname bibitem#1\endcsname}%
\let\auto@bib@innerbib\@empty
\bibitem [{\citenamefont {Epstein}\ \emph {et~al.}(1963)\citenamefont
  {Epstein}, \citenamefont {Goldberger},\ and\ \citenamefont
  {Anfinsen}}]{Epstein1963}%
  \BibitemOpen
  \bibfield  {author} {\bibinfo {author} {\bibfnamefont {C.~J.}\ \bibnamefont
  {Epstein}}, \bibinfo {author} {\bibfnamefont {R.~F.}\ \bibnamefont
  {Goldberger}},\ and\ \bibinfo {author} {\bibfnamefont {C.~B.}\
  \bibnamefont {Anfinsen}},\ }\bibfield  {title} {\bibinfo {title} {The genetic
  control of tertiary protein structure: Studies with model systems},\ }\href
  {https://doi.org/10.1101/SQB.1963.028.01.060} {\bibfield  {journal} {\bibinfo
   {journal} {Cold Spring Harb Symp Quant Biol}\ }\textbf {\bibinfo {volume}
  {28}},\ \bibinfo {pages} {439} (\bibinfo {year} {1963})}\BibitemShut
  {NoStop}%
\bibitem [{\citenamefont {Anfinsen}(1973)}]{Anfinsen1973}%
  \BibitemOpen
  \bibfield  {author} {\bibinfo {author} {\bibfnamefont {C.~B.}\ \bibnamefont
  {Anfinsen}},\ }\bibfield  {title} {\bibinfo {title} {Principles that govern
  the folding of protein chains},\ }\href
  {https://doi.org/10.1126/science.181.4096.223} {\bibfield  {journal}
  {\bibinfo  {journal} {Science}\ }\textbf {\bibinfo {volume} {181}},\ \bibinfo
  {pages} {223} (\bibinfo {year} {1973})}\BibitemShut {NoStop}%
\bibitem [{\citenamefont {G{\=o}}(1983)}]{Go1983}%
  \BibitemOpen
  \bibfield  {author} {\bibinfo {author} {\bibfnamefont {N.}~\bibnamefont
  {G{\=o}}},\ }\bibfield  {title} {\bibinfo {title} {Theoretical studies of
  protein folding},\ }\href
  {https://doi.org/10.1146/annurev.bb.12.060183.001151} {\bibfield  {journal}
  {\bibinfo  {journal} {Ann Rev Biophys Bioeng}\ }\textbf {\bibinfo {volume}
  {12}},\ \bibinfo {pages} {183} (\bibinfo {year} {1983})}\BibitemShut
  {NoStop}%
\bibitem [{\citenamefont {Bryngelson}\ and\ \citenamefont
  {Wolynes}(1987)}]{Bryngelson1987}%
  \BibitemOpen
  \bibfield  {author} {\bibinfo {author} {\bibfnamefont {J.~D.}\ \bibnamefont
  {Bryngelson}}\ and\ \bibinfo {author} {\bibfnamefont {P.~G.}\ \bibnamefont
  {Wolynes}},\ }\bibfield  {title} {\bibinfo {title} {Spin glasses and the
  statistical mechanics of protein folding.},\ }\href
  {https://doi.org/10.1073/pnas.84.21.7524} {\bibfield  {journal} {\bibinfo
  {journal} {P Natl Acad Sci USA}\ }\textbf {\bibinfo {volume} {84}},\ \bibinfo
  {pages} {7524} (\bibinfo {year} {1987})}\BibitemShut {NoStop}%
\bibitem [{\citenamefont {Bryngelson}\ \emph {et~al.}(1995)\citenamefont
  {Bryngelson}, \citenamefont {Onuchic}, \citenamefont {Socci},\ and\
  \citenamefont {Wolynes}}]{Bryngelson1995}%
  \BibitemOpen
  \bibfield  {author} {\bibinfo {author} {\bibfnamefont {J.~D.}\ \bibnamefont
  {Bryngelson}}, \bibinfo {author} {\bibfnamefont {J.~N.}\ \bibnamefont
  {Onuchic}}, \bibinfo {author} {\bibfnamefont {N.~D.}\ \bibnamefont {Socci}},\
  and\ \bibinfo {author} {\bibfnamefont {P.~G.}\ \bibnamefont {Wolynes}},\
  }\bibfield  {title} {\bibinfo {title} {Funnels, pathways, and the energy
  landscape of protein folding: A synthesis},\ }\href
  {https://doi.org/10.1002/prot.340210302} {\bibfield  {journal} {\bibinfo
  {journal} {Proteins}\ }\textbf {\bibinfo {volume} {21}},\ \bibinfo {pages}
  {167} (\bibinfo {year} {1995})}\BibitemShut {NoStop}%
\bibitem [{\citenamefont {Dill}\ and\ \citenamefont {Chan}(1997)}]{Dill1997}%
  \BibitemOpen
  \bibfield  {author} {\bibinfo {author} {\bibfnamefont {K.~A.}\ \bibnamefont
  {Dill}}\ and\ \bibinfo {author} {\bibfnamefont {H.~S.}\ \bibnamefont
  {Chan}},\ }\bibfield  {title} {\bibinfo {title} {From Levinthal to pathways
  to funnels},\ }\href {https://doi.org/10.1038/nsb0197-10} {\bibfield
  {journal} {\bibinfo  {journal} {Nat Struct Biol}\ }\textbf {\bibinfo
  {volume} {4}},\ \bibinfo {pages} {10} (\bibinfo {year}
  {1997})}\BibitemShut {NoStop}%
\bibitem [{\citenamefont {Saito}\ \emph {et~al.}(1997)\citenamefont {Saito},
  \citenamefont {Sasai},\ and\ \citenamefont {Yomo}}]{Saito1997}%
  \BibitemOpen
  \bibfield  {author} {\bibinfo {author} {\bibfnamefont {S.}~\bibnamefont
  {Saito}}, \bibinfo {author} {\bibfnamefont {M.}~\bibnamefont {Sasai}},\ and\
  \bibinfo {author} {\bibfnamefont {T.}~\bibnamefont {Yomo}},\ }\bibfield
  {title} {\bibinfo {title} {Evolution of the folding ability of proteins
  through functional selection},\ }\href
  {https://doi.org/10.1073/pnas.94.21.11324} {\bibfield  {journal} {\bibinfo
  {journal} {P Natl Acad Sci USA}\ }\textbf {\bibinfo {volume} {94}},\ \bibinfo
  {pages} {11324} (\bibinfo {year} {1997})}\BibitemShut {NoStop}%
\bibitem [{\citenamefont {Yomo}\ \emph {et~al.}(1999)\citenamefont {Yomo},
  \citenamefont {Saito},\ and\ \citenamefont {Sasai}}]{Yomo1999}%
  \BibitemOpen
  \bibfield  {author} {\bibinfo {author} {\bibfnamefont {T.}~\bibnamefont
  {Yomo}}, \bibinfo {author} {\bibfnamefont {S.}~\bibnamefont {Saito}},\ and\
  \bibinfo {author} {\bibfnamefont {M.}~\bibnamefont {Sasai}},\ }\bibfield
  {title} {\bibinfo {title} {Gradual development of protein-like global
  structures through functional selection},\ }\href
  {https://doi.org/10.1038/11512} {\bibfield  {journal} {\bibinfo  {journal}
  {Nat Struct Biol}\ }\textbf {\bibinfo {volume} {6}},\ \bibinfo {pages}
  {743} (\bibinfo {year} {1999})}\BibitemShut {NoStop}%
\bibitem [{\citenamefont {Sasaki}\ and\ \citenamefont
  {Sasai}(2002)}]{Sasaki2002}%
  \BibitemOpen
  \bibfield  {author} {\bibinfo {author} {\bibfnamefont {T.}~\bibnamefont
  {Sasaki}}\ and\ \bibinfo {author} {\bibfnamefont {M.}~\bibnamefont {Sasai}},\
  }\bibfield  {title} {\bibinfo {title} {Correlation between the conformation
  space and the sequence space of peptide chain},\ }\href
  {https://doi.org/10.1023/A:1020301714197} {\bibfield  {journal} {\bibinfo
  {journal} {J Biol Phys}\ }\textbf {\bibinfo {volume} {28}},\ \bibinfo {pages}
  {483} (\bibinfo {year} {2002})}\BibitemShut {NoStop}%
\bibitem [{\citenamefont {Nagao}\ \emph {et~al.}(2005)\citenamefont {Nagao},
  \citenamefont {Terada}, \citenamefont {Yomo},\ and\ \citenamefont
  {Sasai}}]{Nagao2005}%
  \BibitemOpen
  \bibfield  {author} {\bibinfo {author} {\bibfnamefont {C.}~\bibnamefont
  {Nagao}}, \bibinfo {author} {\bibfnamefont {T.~P.}\ \bibnamefont {Terada}},
  \bibinfo {author} {\bibfnamefont {T.}~\bibnamefont {Yomo}},\ and\ \bibinfo
  {author} {\bibfnamefont {M.}~\bibnamefont {Sasai}},\ }\bibfield  {title}
  {\bibinfo {title} {Correlation between evolutionary structural development
  and protein folding},\ }\href {https://doi.org/10.1073/pnas.0509163102}
  {\bibfield  {journal} {\bibinfo  {journal} {P Natl Acad Sci USA}\ }\textbf
  {\bibinfo {volume} {102}},\ \bibinfo {pages} {18950} (\bibinfo {year}
  {2005})}\BibitemShut {NoStop}%
\bibitem [{\citenamefont {Sakata}\ \emph {et~al.}(2009)\citenamefont {Sakata},
  \citenamefont {Hukushima},\ and\ \citenamefont {Kaneko}}]{Sakata2009}%
  \BibitemOpen
  \bibfield  {author} {\bibinfo {author} {\bibfnamefont {A.}~\bibnamefont
  {Sakata}}, \bibinfo {author} {\bibfnamefont {K.}~\bibnamefont {Hukushima}},\
  and\ \bibinfo {author} {\bibfnamefont {K.}~\bibnamefont {Kaneko}},\
  }\bibfield  {title} {\bibinfo {title} {Funnel landscape and mutational
  robustness as a result of evolution under thermal noise},\ }\href
  {https://doi.org/10.1103/PhysRevLett.102.148101} {\bibfield  {journal}
  {\bibinfo  {journal} {Phys. Rev. Lett.}\ }\textbf {\bibinfo {volume} {102}},\
  \bibinfo {pages} {148101} (\bibinfo {year} {2009})}\BibitemShut {NoStop}%
\bibitem [{\citenamefont {Riddle}\ \emph {et~al.}(1997)\citenamefont {Riddle},
  \citenamefont {Santiago}, \citenamefont {Bray-Hall}, \citenamefont {Doshi},
  \citenamefont {Grantcharova}, \citenamefont {Yi},\ and\ \citenamefont
  {Baker}}]{Riddle1997}%
  \BibitemOpen
  \bibfield  {author} {\bibinfo {author} {\bibfnamefont {D.~S.}\ \bibnamefont
  {Riddle}}, \bibinfo {author} {\bibfnamefont {J.~V.}\ \bibnamefont
  {Santiago}}, \bibinfo {author} {\bibfnamefont {S.~T.}\ \bibnamefont
  {Bray-Hall}}, \bibinfo {author} {\bibfnamefont {N.}~\bibnamefont {Doshi}},
  \bibinfo {author} {\bibfnamefont {V.~P.}\ \bibnamefont {Grantcharova}},
  \bibinfo {author} {\bibfnamefont {Q.}~\bibnamefont {Yi}},\ and\ \bibinfo
  {author} {\bibfnamefont {D.}~\bibnamefont {Baker}},\ }\bibfield  {title}
  {\bibinfo {title} {Functional rapidly folding proteins from simplified amino
  acid sequences},\ }\href {https://doi.org/10.1038/nsb1097-805} {\bibfield
  {journal} {\bibinfo  {journal} {Nat Struct Biol}\ }\textbf {\bibinfo
  {volume} {4}},\ \bibinfo {pages} {805} (\bibinfo {year} {1997})}\BibitemShut
  {NoStop}%
\bibitem [{\citenamefont {Wang}\ and\ \citenamefont {Wang}(1999)}]{Wang1999}%
  \BibitemOpen
  \bibfield  {author} {\bibinfo {author} {\bibfnamefont {J.}~\bibnamefont
  {Wang}}\ and\ \bibinfo {author} {\bibfnamefont {W.}~\bibnamefont {Wang}},\
  }\bibfield  {title} {\bibinfo {title} {A computational approach to
  simplifying the protein folding alphabet},\ }\href
  {https://doi.org/10.1038/14918} {\bibfield  {journal} {\bibinfo  {journal}
  {Nat Struct Biol}\ }\textbf {\bibinfo {volume} {6}},\ \bibinfo {pages}
  {1033} (\bibinfo {year} {1999})}\BibitemShut {NoStop}%
\bibitem [{\citenamefont {Miyazawa}\ and\ \citenamefont
  {Jernigan}(1999)}]{Miyazawa1999}%
  \BibitemOpen
  \bibfield  {author} {\bibinfo {author} {\bibfnamefont {S.}~\bibnamefont
  {Miyazawa}}\ and\ \bibinfo {author} {\bibfnamefont {R.~L.}\ \bibnamefont
  {Jernigan}},\ }\bibfield  {title} {\bibinfo {title} {Self-consistent
  estimation of inter-residue protein contact energies based on an equilibrium
  mixture approximation of residues},\ }\href
  {https://doi.org/10.1002/(SICI)1097-0134(19990101)34:1<49::AID-PROT5>3.0.CO;2-L}
  {\bibfield  {journal} {\bibinfo  {journal} {Proteins}\ }\textbf {\bibinfo
  {volume} {34}},\ \bibinfo {pages} {49} (\bibinfo {year} {1999})}\BibitemShut
  {NoStop}%
\bibitem [{\citenamefont {Saito}\ and\ \citenamefont
  {Kikuchi}(2013)}]{Saito2013}%
  \BibitemOpen
  \bibfield  {author} {\bibinfo {author} {\bibfnamefont {N.}~\bibnamefont
  {Saito}}\ and\ \bibinfo {author} {\bibfnamefont {M.}~\bibnamefont
  {Kikuchi}},\ }\bibfield  {title} {\bibinfo {title} {Robustness leads close to
  the edge of chaos in coupled map networks: toward the understanding of
  biological networks},\ }\href {https://doi.org/10.1088/1367-2630/15/5/053037}
  {\bibfield  {journal} {\bibinfo  {journal} {New J Phys}\ }\textbf {\bibinfo
  {volume} {15}},\ \bibinfo {pages} {053037} (\bibinfo {year}
  {2013})}\BibitemShut {NoStop}%
\bibitem [{\citenamefont {Nagata}\ and\ \citenamefont
  {Kikuchi}(2020)}]{Nagata2020}%
  \BibitemOpen
  \bibfield  {author} {\bibinfo {author} {\bibfnamefont {S.}~\bibnamefont
  {Nagata}}\ and\ \bibinfo {author} {\bibfnamefont {M.}~\bibnamefont
  {Kikuchi}},\ }\bibfield  {title} {\bibinfo {title} {Emergence of cooperative
  bistability and robustness of gene regulatory networks},\ }\href
  {https://doi.org/10.1371/journal.pcbi.1007969} {\bibfield  {journal}
  {\bibinfo  {journal} {PLoS Comput Biol}\ }\textbf {\bibinfo {volume} {16}},\
  \bibinfo {pages} {e1007969} (\bibinfo {year} {2020})}\BibitemShut {NoStop}%
\bibitem [{\citenamefont {Kaneko}\ and\ \citenamefont
  {Kikuchi}(2022)}]{Kaneko2022}%
  \BibitemOpen
  \bibfield  {author} {\bibinfo {author} {\bibfnamefont {T.}~\bibnamefont
  {Kaneko}}\ and\ \bibinfo {author} {\bibfnamefont {M.}~\bibnamefont
  {Kikuchi}},\ }\bibfield  {title} {\bibinfo {title} {Evolution enhances
  mutational robustness and suppresses the emergence of a new phenotype: A new
  computational approach for studying evolution},\ }\href
  {https://doi.org/10.1371/journal.pcbi.1009796} {\bibfield  {journal}
  {\bibinfo  {journal} {PLoS Comput Biol}\ }\textbf {\bibinfo {volume} {18}},\
  \bibinfo {pages} {e1009796} (\bibinfo {year} {2022})}\BibitemShut {NoStop}%
\bibitem [{\citenamefont {Kikuchi}(2024)}]{Kikuchi2024}%
  \BibitemOpen
  \bibfield  {author} {\bibinfo {author} {\bibfnamefont {M.}~\bibnamefont
  {Kikuchi}},\ }\bibfield  {title} {\bibinfo {title} {Phenotype selection due
  to mutational robustness},\ }\href
  {https://doi.org/10.1371/journal.pone.0311058} {\bibfield  {journal}
  {\bibinfo  {journal} {PLoS ONE}\ }\textbf {\bibinfo {volume} {19}},\ \bibinfo
  {pages} {e0311058} (\bibinfo {year} {2024})}\BibitemShut {NoStop}%
\bibitem [{\citenamefont {Berg}\ and\ \citenamefont
  {Neuhaus}(1992)}]{Berg1992a}%
  \BibitemOpen
  \bibfield  {author} {\bibinfo {author} {\bibfnamefont {B.~A.}\ \bibnamefont
  {Berg}}\ and\ \bibinfo {author} {\bibfnamefont {T.}~\bibnamefont {Neuhaus}},\
  }\bibfield  {title} {\bibinfo {title} {Multicanonical ensemble: A new
  approach to simulate first-order phase transitions},\ }\href
  {https://doi.org/10.1103/PhysRevLett.68.9} {\bibfield  {journal} {\bibinfo
  {journal} {Phys. Rev. Lett.}\ }\textbf {\bibinfo {volume} {68}},\ \bibinfo
  {pages} {9} (\bibinfo {year} {1992})}\BibitemShut {NoStop}%
\bibitem [{\citenamefont {Berg}\ and\ \citenamefont {Celik}(1992)}]{Berg1992b}%
  \BibitemOpen
  \bibfield  {author} {\bibinfo {author} {\bibfnamefont {B.~A.}\ \bibnamefont
  {Berg}}\ and\ \bibinfo {author} {\bibfnamefont {T.}~\bibnamefont {Celik}},\
  }\bibfield  {title} {\bibinfo {title} {New approach to spin-glass
  simulations},\ }\href {https://doi.org/10.1103/PhysRevLett.69.2292}
  {\bibfield  {journal} {\bibinfo  {journal} {Phys. Rev. Lett.}\ }\textbf
  {\bibinfo {volume} {69}},\ \bibinfo {pages} {2292} (\bibinfo {year}
  {1992})}\BibitemShut {NoStop}%
\bibitem [{\citenamefont {Lee}(1993)}]{Lee1993}%
  \BibitemOpen
  \bibfield  {author} {\bibinfo {author} {\bibfnamefont {J.}~\bibnamefont
  {Lee}},\ }\bibfield  {title} {\bibinfo {title} {New Monte Carlo algorithm:
  Entropic sampling},\ }\href {https://doi.org/10.1103/PhysRevLett.71.211}
  {\bibfield  {journal} {\bibinfo  {journal} {Phys. Rev. Lett.}\ }\textbf
  {\bibinfo {volume} {71}},\ \bibinfo {pages} {211} (\bibinfo {year}
  {1993})}\BibitemShut {NoStop}%
\bibitem [{\citenamefont {Wang}\ and\ \citenamefont
  {Landau}(2001{\natexlab{a}})}]{Wang2001a}%
  \BibitemOpen
  \bibfield  {author} {\bibinfo {author} {\bibfnamefont {F.}~\bibnamefont
  {Wang}}\ and\ \bibinfo {author} {\bibfnamefont {D.~P.}\ \bibnamefont
  {Landau}},\ }\bibfield  {title} {\bibinfo {title} {Efficient, multiple-range
  random walk algorithm to calculate the density of states},\ }\href
  {https://doi.org/10.1103/PhysRevLett.86.2050} {\bibfield  {journal} {\bibinfo
   {journal} {Phys. Rev. Lett.}\ }\textbf {\bibinfo {volume} {86}},\ \bibinfo
  {pages} {2050} (\bibinfo {year} {2001}{\natexlab{a}})}\BibitemShut {NoStop}%
\bibitem [{\citenamefont {Wang}\ and\ \citenamefont
  {Landau}(2001{\natexlab{b}})}]{Wang2001b}%
  \BibitemOpen
  \bibfield  {author} {\bibinfo {author} {\bibfnamefont {F.}~\bibnamefont
  {Wang}}\ and\ \bibinfo {author} {\bibfnamefont {D.~P.}\ \bibnamefont
  {Landau}},\ }\bibfield  {title} {\bibinfo {title} {Determining the density of
  states for classical statistical models: A random walk algorithm to produce a
  flat histogram},\ }\href {https://doi.org/10.1103/PhysRevE.64.056101}
  {\bibfield  {journal} {\bibinfo  {journal} {Phys. Rev. E}\ }\textbf {\bibinfo
  {volume} {64}},\ \bibinfo {pages} {056101} (\bibinfo {year}
  {2001}{\natexlab{b}})}\BibitemShut {NoStop}%
\bibitem [{\citenamefont {Srere}(1984)}]{Srere1984}%
  \BibitemOpen
  \bibfield {author} {\bibinfo {author} {\bibfnamefont {P.~A.}\ \bibnamefont
  {Srere}},\ }\bibfield {title} {\bibinfo {title} {Why are enzymes so big?},\ }
  \href {https://doi.org/10.1016/0968-0004(84)90221-4} {\bibfield {journal}
  {\bibinfo {journal} {Trends Biochem. Sci.}\ }\textbf {\bibinfo {volume}
  {9}},\ \bibinfo {pages} {387} (\bibinfo {year} {1984})}\BibitemShut {NoStop}%
\bibitem [{\citenamefont {Chothia}(1992)}]{Chothia1992}%
  \BibitemOpen
  \bibfield {author} {\bibinfo {author} {\bibfnamefont {C.}\ \bibnamefont
  {Chothia}},\ }\bibfield {title} {\bibinfo {title} {One thousand families for
  the molecular biologist},\ }\href {https://doi.org/10.1038/357543a0}
  {\bibfield {journal} {\bibinfo {journal} {Nature}\ }\textbf {\bibinfo
  {volume} {357}},\ \bibinfo {pages} {543} (\bibinfo {year}
  {1992})}\BibitemShut {NoStop}%
\bibitem [{\citenamefont {Deutsch}\ and\ \citenamefont
  {Kurosky}(1996)}]{Deutch1996}%
  \BibitemOpen
  \bibfield  {author} {\bibinfo {author} {\bibfnamefont {J.~M.}\ \bibnamefont
  {Deutsch}}\ and\ \bibinfo {author} {\bibfnamefont {T.}~\bibnamefont
  {Kurosky}},\ }\bibfield  {title} {\bibinfo {title} {New algorithm for protein
  design},\ }\href {https://doi.org/10.1103/PhysRevLett.76.323} {\bibfield
  {journal} {\bibinfo  {journal} {Phys. Rev. Lett.}\ }\textbf {\bibinfo
  {volume} {76}},\ \bibinfo {pages} {323} (\bibinfo {year} {1996})}\BibitemShut
  {NoStop}%
\bibitem [{\citenamefont {Iba}\ \emph {et~al.}(1998)\citenamefont {Iba},
  \citenamefont {Tokita},\ and\ \citenamefont {Kikuch}}]{Iba1998}%
  \BibitemOpen
  \bibfield  {author} {\bibinfo {author} {\bibfnamefont {Y.}~\bibnamefont
  {Iba}}, \bibinfo {author} {\bibfnamefont {K.}~\bibnamefont {Tokita}},\ and\
  \bibinfo {author} {\bibfnamefont {M.}~\bibnamefont {Kikuchi}},\ }\bibfield
  {title} {\bibinfo {title} {Design equation: A novel approach to heteropolymer
  design},\ }\href {https://doi.org/10.1143/JPSJ.67.3985} {\bibfield  {journal}
  {\bibinfo  {journal} {J Phys Soc Jpn}\ }\textbf {\bibinfo {volume} {67}},\
  \bibinfo {pages} {3985} (\bibinfo {year} {1998})}\BibitemShut {NoStop}%
\bibitem [{\citenamefont {Takahashi}\ \emph {et~al.}(2021)\citenamefont
  {Takahashi}, \citenamefont {Chikenji},\ and\ \citenamefont
  {Tokita}}]{Takahashi2021}%
  \BibitemOpen
  \bibfield  {author} {\bibinfo {author} {\bibfnamefont {T.}~\bibnamefont
  {Takahashi}}, \bibinfo {author} {\bibfnamefont {G.}~\bibnamefont
  {Chikenji}},\ and\ \bibinfo {author} {\bibfnamefont {K.}~\bibnamefont
  {Tokita}},\ }\bibfield  {title} {\bibinfo {title} {Lattice protein design
  using Bayesian learning},\ }\href
  {https://doi.org/10.1103/PhysRevE.104.014404} {\bibfield  {journal} {\bibinfo
  {journal} {Phys. Rev. E}\ }\textbf {\bibinfo {volume} {104}},\ \bibinfo
  {pages} {014404} (\bibinfo {year} {2021})}\BibitemShut {NoStop}%
\bibitem [{\citenamefont {Wolynes}(1997)}]{Wolynes1997}%
  \BibitemOpen
  \bibfield  {author} {\bibinfo {author} {\bibfnamefont {P.~G.}\ \bibnamefont
  {Wolynes}},\ }\bibfield  {title} {\bibinfo {title} {As simple as can be?},\
  }\href {https://doi.org/10.1038/nsb1197-871} {\bibfield  {journal} {\bibinfo
  {journal} {Nat Struct Biol}\ }\textbf {\bibinfo {volume} {4}},\ \bibinfo
  {pages} {871} (\bibinfo {year} {1997})}\BibitemShut {NoStop}%
\end{thebibliography}
%
\section*{Choice of the amino-acid alphabet}

The four-letter alphabet avoids the extensive accidental degeneracy of the
standard two-letter hydrophobic--polar (HP) lattice model.  For a 20-residue HP
chain, exhaustive enumeration reveals many geometrically distinct conformations
with identical energy, making funnel-like energetic organization difficult to
resolve; related limitations of small alphabets have been noted
previously~\cite{Wolynes1997}.  Experimental and computational studies have
shown that a five-letter alphabet with
three hydrophobic and two oppositely charged polar types can retain essential
folding properties~\cite{Riddle1997,Wang1999}.  Guided by these results, we
adopt a slightly simplified four-letter alphabet with two hydrophobic and two
oppositely charged polar types.  The contact energies were chosen with
reference to the Miyazawa--Jernigan potential~\cite{Miyazawa1999} and adjusted
to suppress accidental degeneracy.  No contact-energy parameter was optimized
during sequence sampling or evolution.

\section*{Conformational ensemble}

Repeated equilibrium evaluations of fitness dominate the computational cost of
sequence-space sampling.  We therefore restricted the sums in
Eq.~(\ref{eq:fitness}) to all conformations with 9--12 noncovalent contacts.
These compact conformations contain the relevant low-energy states, whereas
the omitted, less compact conformations contribute predominantly at higher
energies.  To assess the effect of this restriction, we enumerated the complete
conformational ensemble for the highest-fitness sequence at $T=1.0$ and
recomputed its energy spectrum and free-energy landscapes.  The restricted and
complete ensembles yield nearly indistinguishable results, with the native and
denatured basins and the intervening barrier preserved.  A comparison at
$T=1.0$ and at their respective folding temperatures is shown in Supplemental
Material.

\section*{Sequence-space multicanonical sampling}

At each environmental temperature, we performed multicanonical Monte Carlo
sampling~\cite{Berg1992a,Berg1992b} in sequence space, using fitness $f$ as the
sampling variable so that sequences were sampled approximately uniformly
across the accessible fitness range~\cite{Saito2013,Nagata2020,Kaneko2022,
Kikuchi2024}.  The interval $0\le f\le1$ was divided into 100 bins.  A
preliminary Wang--Landau simulation~\cite{Wang2001a,Wang2001b} estimated the
multicanonical weight.  With this weight held fixed, production sampling was
performed using the entropic-sampling method~\cite{Lee1993}.  Because
temperature enters the definition of fitness in Eq.~(\ref{eq:fitness}), an
independent weight was determined for each $T$.

We define one Monte Carlo step (MCS) as 20 attempted single-residue updates,
each applied to a randomly selected residue; thus, each residue is selected
once per MCS on average.  One sequence was recorded after each MCS.  The length
of each production run was chosen to yield approximately 4000 recorded
sequences per accessible fitness bin.  At $T=1.0$, 76 bins were accessible,
and each run comprised $4000\times76$ MCS; at $T=0.1$, all 100 bins were
accessible, and each run comprised $4000\times100$ MCS.  Four independent runs
were initiated from different sequences at each temperature.  Their fitness
distributions and measured observables were statistically consistent, so the
four samples were combined for analysis.

At $T=1.0$, the landscape and fold-class analyses use sequences from the
extreme high-fitness tail sampled by this procedure.  Because evolution
proceeds through stochastic mutation and selection, it need not attain the
mathematical fitness maximum.  At $T=0.1$, the nearly constant genotypic entropy
shows that sufficiently high-fitness sequences remain abundant.  We therefore
represent this physically relevant regime by the highest-fitness sequences
with $f<0.9$.  The extreme tail near $f=1$ is analyzed separately in
Supplemental Material.

\clearpage

\begin{flushleft}{\large Supplemental Material for ``Funnel-like protein energy landscapes
emerge from functional evolution under thermal fluctuations}
\end{flushleft}
\vspace*{1cm}

\renewcommand{\thefigure}{S\arabic{figure}}

\section{Complete conformational ensemble and the folding temperature}

\begin{figure}[htb]
\centering
\includegraphics[width=0.62\columnwidth]{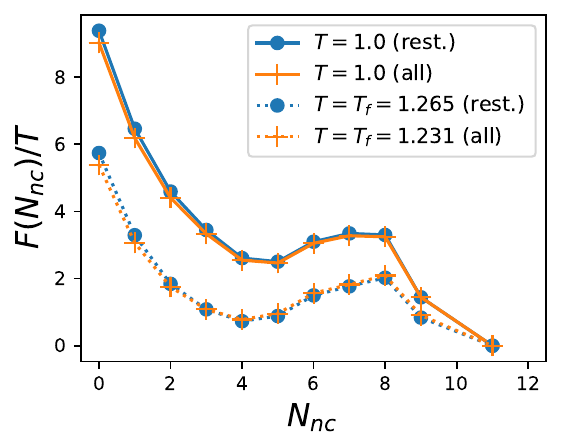}
\caption{Dimensionless free-energy landscapes $F(N_{\rm nc})/T$ of the
highest-fitness sequence obtained at environmental temperature $T=1.0$, using
the restricted 9--12-contact ensemble and the complete conformational ensemble.
Solid lines show the landscapes at $T=1.0$, and dotted lines show those at the
respective folding temperatures, $T_f=1.265$ for the restricted ensemble and
$T_f=1.231$ for the complete ensemble.  Blue circles and orange plus signs
denote the restricted and complete ensembles, respectively.  Each landscape
is shifted so that its minimum is zero.  No conformation exists at
$N_{\rm nc}=10$.}
\label{FigS1}
\end{figure}

During sequence-space sampling, fitness was evaluated using conformations with
9--12 noncovalent contacts.  We assessed the effect of this restriction for the
highest-fitness sequence obtained at environmental temperature $T=1.0$ by
recomputing its free-energy landscape with the complete conformational
ensemble.  Figure~\ref{FigS1} compares the restricted and complete results at
$T=1.0$ and at their respective folding temperatures.

For this sequence, we operationally define $N_{\rm nc}=0$--5 as the denatured
basin, $N_{\rm nc}=9$ and 11 as the native basin, and $N_{\rm nc}=6$--8 as the
intervening barrier region, based on the visible separation of the two basins.
Let $P(n;T)$ denote the equilibrium probability of $N_{\rm nc}=n$.  We define
the folding temperature by
\[
\begin{aligned}
P_D(T)&=\sum_{n=0}^{5}P(n;T),\\
P_N(T)&=P(9;T)+P(11;T),\\
P_D(T_f)&=P_N(T_f).
\end{aligned}
\]
This gives $T_f=1.265$ for the restricted ensemble and $T_f=1.231$ for the
complete ensemble.  The lower value for the complete ensemble is expected
because the additional less compact, predominantly high-energy conformations
increase the statistical weight of the denatured basin.  At $T=1.0$ and at
their respective folding temperatures, the restricted and complete
free-energy landscapes are nearly indistinguishable [Fig.~\ref{FigS1}].  Their
largest difference occurs in the least compact region and is negligible over
the native and denatured basins and the intervening barrier.  Both display
denatured and native minima separated by a barrier.  Thus, for this sequence,
the conformational restriction slightly shifts the folding temperature but
leaves the two-state free-energy organization essentially unchanged.

\section{Native-structure classes and associated free-energy landscapes}

We classified the native conformations of the 1000 highest-fitness sequences
sampled at $T=1.0$.  The sequences form 500 pairs related by reversal of the
chain direction.  After identifying each reversal pair as a single sequence
pattern, the 500 patterns fall into only four native-structure classes:
(a) 243, (b) 170, (c) 25, and (d) 62 patterns.  The pairing of every sequence
with its chain-reversed counterpart and the reproducibility among independent
runs indicate that the highest-fitness region was sampled thoroughly.

We further examined how these native-structure classes are related to the
free-energy landscapes.  The same 200 highest-fitness sequences used for the
landscape survey were regrouped by native-structure class.  For every sequence,
the dimensionless free energy $F(N_{\rm nc})/T$ was shifted so that its minimum
is zero.

\begin{figure*}[htb]
\centering
\includegraphics[width=\textwidth]{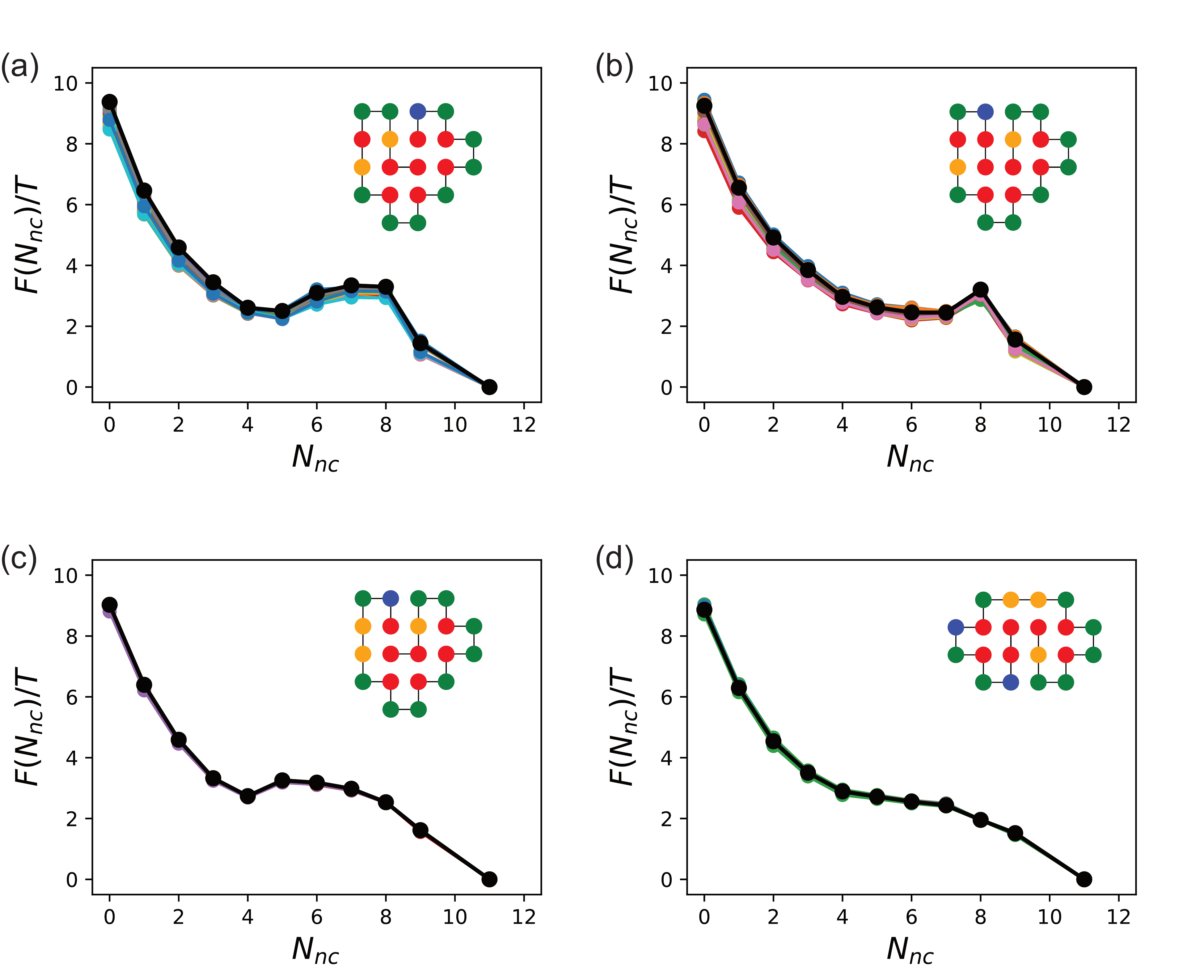}
\caption{Native-structure classes and their associated dimensionless
free-energy landscapes at $T=1.0$.  Panels (a)--(d) show $F(N_{\rm nc})/T$
for all members of the corresponding class among the 200 highest-fitness
sequences used for the landscape survey.  The black profile in each panel is
the highest-fitness member of that class.  The inset shows its native
conformation, with residues colored according to the sequence; red and orange
circles denote hydrophobic residues, and green and blue circles denote polar
residues.  The same four sequences are examined at different temperatures in
Fig.~\ref{FigS3}.  Among the 500 sequence patterns, the classes contain
(a) 243, (b) 170, (c) 25, and (d) 62 patterns.  Each profile is shifted so that
its minimum is zero.}
\label{FigS2}
\end{figure*}

The first three classes share essentially the same overall fold and differ by
a shift of chain registration along the backbone.  The fourth class is
geometrically distinct and invariant under chain reversal.  The native
conformations of all sequences in the four classes possess recognizable
hydrophobic cores, with most hydrophobic residues in the interior and polar
residues preferentially exposed.  Thus the same local functional requirement
is supported by only a small set of global structures within this model.

The variation among the $T=1.0$ free-energy landscapes is closely associated
with native structure [Fig.~\ref{FigS2}].  Profiles of sequences belonging to
the same class nearly coincide, whereas the characteristic shape differs among
classes.  Classes (a)--(c) exhibit clear two-basin organization at $T=1.0$,
with an intermediate-$N_{\rm nc}$ barrier separating the native and denatured
regions.  For class (d), the barrier at $T=1.0$ is less pronounced.

\begin{figure*}[htb]
\centering
\includegraphics[width=\textwidth]{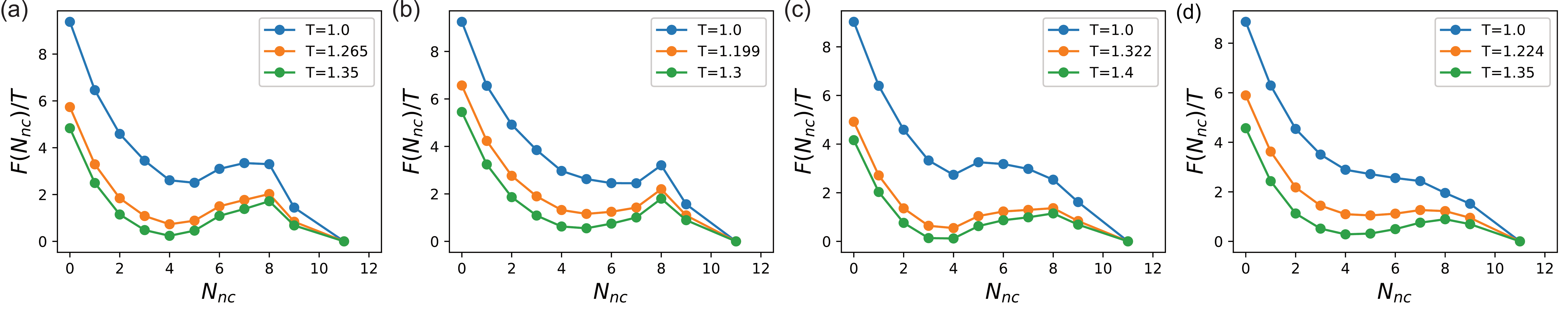}
\caption{Temperature dependence of the dimensionless free-energy landscapes
$F(N_{\rm nc})/T$ of the highest-fitness members of the four native-structure
classes in Fig.~\ref{FigS2}.  Panels (a)--(d) correspond to the same classes as
in Fig.~\ref{FigS2}.  Blue curves show $T=1.0$; orange curves show the
class-specific folding temperatures $T_f=1.265$, 1.199, 1.322, and 1.224; and
green curves show $T=1.35$, 1.3, 1.4, and 1.35, respectively.  The green curves
therefore represent temperatures slightly above the corresponding $T_f$.
Each landscape is shifted so that its minimum is zero.}
\label{FigS3}
\end{figure*}

To examine this organization near coexistence, we determined the folding
temperature of the highest-fitness member of each class, shown by the black
profiles in Fig.~\ref{FigS2}.  For classes (a)--(d), the denatured ensemble was
operationally defined as $N_{\rm nc}=0$--5, 0--7, 0--4, and 0--6,
respectively, based on the visible separation of the low-$N_{\rm nc}$ region
from the native basin.  The native ensemble was defined as $N_{\rm nc}=9$ and
11 for all four classes.  Applying the equal-probability condition
$P_D(T_f)=P_N(T_f)$ gives $T_f=1.265$, 1.199, 1.322, and 1.224 for classes
(a)--(d), respectively.  At these folding temperatures, all four landscapes
display denatured and native basins separated by an intervening barrier
[Fig.~\ref{FigS3}].  In particular, the two-basin organization of class (d),
whose barrier is less pronounced at $T=1.0$, is clearly resolved near $T_f$.
At the slightly higher temperatures also shown in Fig.~\ref{FigS3}, the
two-basin structure persists, with the denatured basin having the larger
equilibrium probability, as expected for $T>T_f$.

\section{Low-temperature landscape and structural diversity}

To characterize the heterogeneous low-temperature regime beyond the
representative sequences shown in the main text, we examined two sets of 200
sequences: the highest-fitness sequences with $f<0.9$ and those in the extreme
$f\simeq1$ tail.  Within each set, sequences were ordered by decreasing fitness
and divided without overlap among panels (a)--(d), which show ranks 1--50,
51--100, 101--150, and 151--200, respectively.  For every sequence,
$F(N_{\rm nc})/T$ was shifted so that its minimum is zero.

\begin{figure*}[htb]
\includegraphics[width=\textwidth]{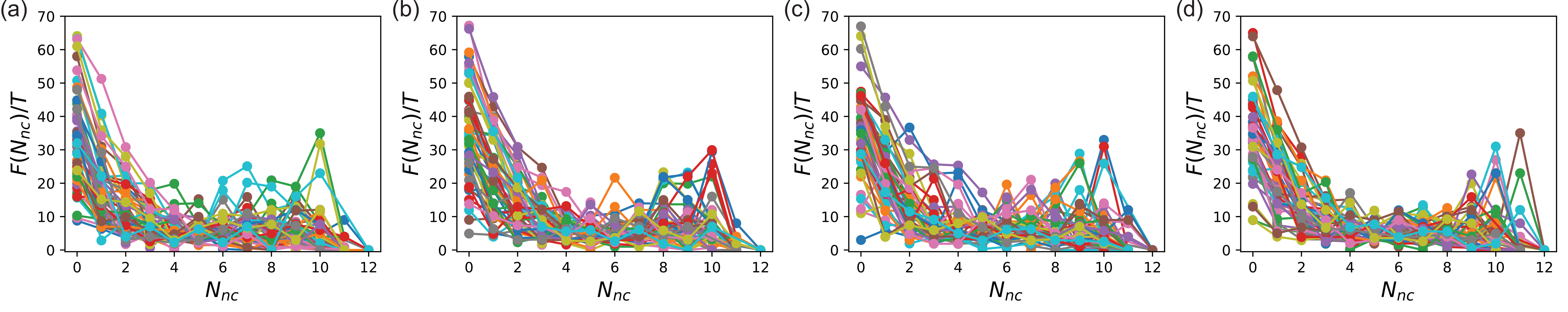}
\caption{Dimensionless free-energy landscapes $F(N_{\rm nc})/T$ at $T=0.1$
for the 200 highest-fitness sequences with $f<0.9$.  Panels (a)--(d)
show successive groups of 50 sequences in decreasing order of fitness.}
\label{FigS4}
\end{figure*}

\begin{figure*}[htb]
\includegraphics[width=\textwidth]{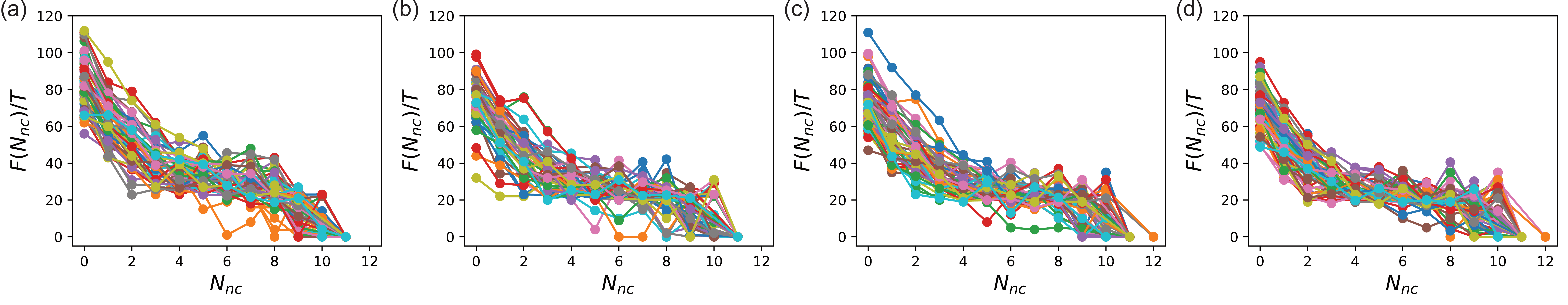}
\caption{Dimensionless free-energy landscapes $F(N_{\rm nc})/T$ of the 200
highest-fitness sequences sampled at $T=0.1$, representing the extreme
$f\simeq1$ tail.  Panels (a)--(d) show successive groups of 50 sequences in
decreasing order of fitness.}
\label{FigS5}
\end{figure*}

In the broadly populated high-fitness regime represented by the $f<0.9$ set,
the landscapes are strongly rugged and show no common funnel-like organization
[Fig.~\ref{FigS4}].  Even in the extreme highest-fitness tail, an overall
native-directed bias emerges but substantial roughness and sequence-to-sequence
variation remain [Fig.~\ref{FigS5}].

The structural diversity is likewise much greater than at $T=1.0$.  The same
chain-reversal identification was applied to both samples.  After
chain-reversed pairs were identified, the 200 sequences in the extreme
$f\simeq1$ tail span 107 distinct native conformations.  In the $f<0.9$ set,
the same identification removes 13 sequences, leaving 187 sequence patterns,
each with a distinct native conformation.  Both samples thus show that, at
$T=0.1$, high fitness is realized by a wide diversity of native conformations
and the local functional motif does not appreciably constrain global structure.

\section{Population-based evolutionary simulations}

We tested whether the sequences identified by multicanonical sampling are
accessible through explicit evolutionary simulations.  A population of 1000 random
sequences was evolved at $T=1.0$.  Each individual of fitness $f$ was assigned
a stochastic selection score

\begin{equation}
w=r\exp(f/Q),
\end{equation}

\noindent where $r$ is uniformly distributed on $[0,1)$ and $Q=2$.  The 500 individuals
with the largest scores survived, each produced one offspring, and every
offspring underwent one mutation in which a randomly chosen residue was
replaced by one of the other three residue types.  This restored the population
to 1000 sequences each generation.

\begin{figure}[htb]
\includegraphics[width=\columnwidth]{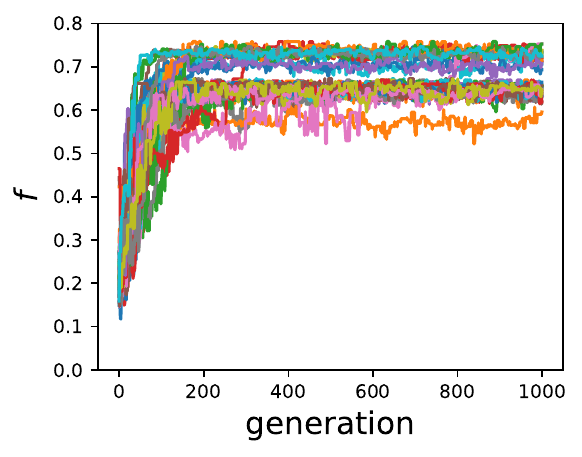}
\caption{Maximum fitness in the population versus generation for 40 independent
evolutionary simulations at $T=1.0$.}
\label{FigS6}
\end{figure}

We performed 40 independent simulations of 1000 generations.  As shown in
Fig.~\ref{FigS6}, maximum fitness rose rapidly and then approached one of
several plateau values.  In two runs, the same highest-fitness sequence
identified by multicanonical sampling appeared during the evolutionary
trajectory.  Its appearance demonstrates that the sequence is not an artifact
of multicanonical sampling but is reachable under the specified point-mutation
and selection dynamics.  The lower
plateaus in the remaining runs reflect the limited exploration produced by
single-residue mutations on the simulated time scale.

\end{document}